\documentclass[12pt,leqno]{elsart}

\usepackage{amsfonts,amsmath,amssymb,algorithm,epsfig,fancyhdr,graphicx,mathrsfs,eucal}
\usepackage{algpseudocode}
\usepackage{flafter}
\usepackage{mathrsfs}
\usepackage{natbib}
\usepackage{url}

\def\BibTeX{{\rm B\kern-.05em{\sc i\kern-.025em b}\kern-.08em
    T\kern-.1667em\lower.7ex\hbox{E}\kern-.125emX}}

\numberwithin{equation}{section}

\newtheorem{theorem}{Theorem}

\newtheorem{corollary}[theorem]{Corollary}

\newtheorem{lemma}[theorem]{Lemma}

\newtheorem{problem}[theorem]{Problem}
\newtheorem{proposition}[theorem]{Proposition}
\newtheorem{remark}[theorem]{Remark}

\newenvironment{proof}[1][Proof]{\textbf{#1.} }{\ \rule{0.5em}{0.5em}}

\renewcommand\emptyset{\varnothing}

\newcommand\aff{\operatorname{aff}}
\newcommand\Arcs{\operatorname{Arcs}}

\newcommand\conv{\operatorname{conv}}

\newcommand\inter{\operatorname{int}}

\newcommand\rel{\operatorname{rel}}
\newcommand\rank{\operatorname{rank}}
\newcommand\Span{\operatorname{span}}
\newcommand\sgn{\operatorname{sgn}} \newcommand\Star{\operatorname{Star}}

\newcommand\Test{\operatorname{Test}} \newcommand\Update{\operatorname{Update}}

\renewcommand{\phi}{\varphi}                 
\renewcommand{\epsilon}{\varepsilon}

\newcommand{\aaa}{\mbox{\boldmath$a$}}
\newcommand{\bb}{\mbox{\boldmath$b$}}
\newcommand{\cc}{\mbox{\boldmath$c$}}
\newcommand{\dd}{\mbox{\boldmath$d$}}
\renewcommand{\e}{\mbox{\boldmath$e$}}

\newcommand{\n}{\mbox{\boldmath$n$}}

\newcommand{\p}{\mbox{\boldmath$p$}}

\newcommand{\vv}{\mbox{\boldmath$v$}}
\newcommand{\uu}{\mbox{\boldmath$u$}}
\newcommand{\w}{\mbox{\boldmath$w$}}
\newcommand{\x}{\mbox{\boldmath$x$}}
\newcommand{\y}{\mbox{\boldmath$y$}}

\newcommand\B{\mathbf{B}}
\newcommand\bS{\mathbf{S}}

\newcommand\fc{c}
\newcommand\ff{f}

\newcommand\fp{p}
\newcommand\fq{q}

\newcommand\R{\mathbb{R}}
\newcommand\HH{\mathbb{H}}
\newcommand\SSS{\mathbb{S}}
\newcommand\X{\mathbb{X}}
\newcommand\Z{\mathbb{Z}}

\newcommand\cC{\mathcal{C}}
\newcommand\cF{\mathscr{F}}

\newcommand\cI{\mathscr{I}}
\newcommand\cR{\mbox{{\slshape\textsf{R}}}}
\newcommand\cM{\mathscr{M}}
\newcommand\cN{\mathscr{N}}
\newcommand\cP{\mathcal{P}}
\newcommand\cS{\mathcal{S}}
\newcommand\cU{\mathscr{U}}

\newcommand\ssf{\textsf{f}}

\begin{document}
\begin{frontmatter}
\title
{An Efficient Local Approach to Convexity Testing of Piecewise-Linear
Hypersurfaces}
\author{Konstantin Rybnikov}

\address{One University Ave., Olney Hall 428, University of Massachusetts at Lowell, Lowell, MA 01854 USA}
\address{Email address: \emph{\texttt{krybniko@cs.uml.edu}}}

\begin{abstract}
We show that a closed piecewise-linear hypersurface immersed in $\R^n$ ($n\ge 3$) is the boundary of a convex body
 if and only if  every point in the interior of each $(n-3)$-face has a neighborhood that lies on the boundary of some convex body; no assumptions about the hypersurface's topology are needed. We  derive this criterion from our
generalization of Van Heijenoort's (1952) theorem on locally convex hypersurfaces in $\R^n$ to spherical spaces. We also give  an easy-to-implement convexity testing algorithm, which is based on our criterion.
 For $\R^3$ the number of arithmetic operations used by the algorithm  is at most linear in the number of vertices,
while in general it is at most linear in the number of incidences between the $(n-2)$-faces and
$(n-3)$-faces. When the dimension $n$ is not fixed and only ring arithmetic is allowed,
 the algorithm still remains  polynomial. Our method
works in more general situations  than the convexity verification algorithms
developed by \citet{M96b} and \citet{Dev} -- for example, our method does not require the
input surface to be orientable, nor it requires the input data to include   normal vectors to the facets
that are oriented ``in a coherent way".   For $\R^3$ the complexity of our algorithm is the
same as that of previous algorithms; for higher dimensions there seems to be no clear
winner, but our approach is the only one that easily handles inputs in which the facet normals
are not known to be coherently oriented or are not given at all. Furthermore, our method can be extended to piecewise-polynomial surfaces of
small degree.
\end{abstract}
\end{frontmatter}
Time 00:28 \quad \today

\section{Introduction} \vspace{-2mm} Blum and Kannan (1989) suggested a
paradigm of output verification.   Since a complete check of a program is often
difficult or not possible -- for example, when the source code has not been made public -- it is
important to have algorithms that verify key properties of mathematical objects
generated by programs. Instead of the source code verification one can try to verify the
properties of the output that are deemed essential by users of the program. In
computational geometry this paradigm was developed, among others, by Melhorn et al.
 (1996, 1999) and Devillers et al. (1998). For example, the LEDA C++ library  contains
programs verifying convexity of a polygon, Delaunay property of a tiling, etc (Mehlhorn \and N\"{a}her, 2000).  Devillers et al.
(1998) argue that it is easier to evaluate the quality of the output of a geometric
algorithm, than the correctness of the algorithm or program producing it. This paper
contributes to the problem of verification of convexity of a large class of
piecewise-linear (PL)  hypersurfaces in $\R^n$ for $n\ge3$. The novelty of our approach
is in reducing the verification of global convexity of a PL-hypersurface to
the verification of local convexity at the faces of small codimension.  We show that a
closed bounded PL-hypersurface  realized in $\R^n$ ($n\ge3$) without local
self-intersections
  is the boundary of a convex body if and only if
each $(n-3)$-face of the hypersurface has a point in its relative interior such that a
small Euclidean ball at this point is cut by the hypersurface into two pieces, one of
which is convex.  The local convexity condition can also be expressed as that the point
has a neighborhood on the hypersurface, which lies on the boundary of a convex body.
 If the hypersurface
is not bounded, this criterion is invalid. However, \emph{if} it is known that  the
hypersurface has \emph{at least one point of strict convexity} -- i.e., a point $s$
such that a small ball, centered at $s$, intersects the hypersurface over a set $S$,
which lies on the boundary of a convex body and such that $S\: \diagdown\: s$  lies in
an \emph{open halfspace} with respect to some hyperplane through $s$ -- our test can
still be used. In fact, our result is slightly more general: we prove our criterion for
any cellwise-flat PL-realization in $\R^n$ ($n \ge 3$) of a semiregular CW-partition of
an $(n-1)$-manifold. The technical terms used in this formulation are defined in the
next section.

In this paper we also construct an algorithm for convexity testing that can be applied
to any closed  PL-hypersurface. Our approach does not require any preliminary knowledge
of the topology of the surface, or any information  about its orientability.
The direct comparison of the complexity of our algorithm and those of Mehlhorn
et al. (1996b; 1999) and Devillers et al. (1998)  is not quite meaningful, since these
authors make the following simplifying assumptions.
\begin{enumerate}
\item[(CO)] \emph{The input is known to be an orientable closed hypersurface. The normals to the $(n-1)$-faces are given as part of the input, and they are all
oriented   either outwards or inwards (``coherent orientation")}
\item[(S)]  \emph{the cell-partition is simplicial}.
\end{enumerate}

The last condition is  not needed for these algorithms to work; however, it does affect
the complexity analysis, which is given in the papers only for the simplicial case. It
is not clear to us why the assumption (CO) is natural; in any
case, they significantly simplify all considerations. Our approach
does not require the assumption (CO). For $\R^3$, if  conditions (CO) and (C) are met,  both our and previous approaches have the same complexity, which is big-$O$ in the number of vertices. While Devillers et al.(1998) claimed $O(f_0)$  as the running time of for any dimension,  this was clearly a typo: first, this bound is  impossible, even for PL-spheres, by fundamental counting theorems of polyhedral combinatorics and second, the pseudocode in (Devillers et al, 1998: Section 3) has running time, in the notation of that paper, of $\underset{j=0}{\overset{d-2}{\sum}} \ssf_j+ \underset{\{F \in \Gamma\;|\;\dim F=d-2
\}}{\sum} \underset{{j=3}}{\overset{d}{\sum}} \ssf_{j-1,j}(F)$.

Formal definitions and notation are given in Section \ref{sec:definitions}. The paper
presupposes some familiarity with partially ordered sets (posets), linear algebra and
geometry \citep[e.g.][]{KM}, basic convexity theory \citep[e.g.][]{Rock}, general
topology \citep[e.g.][]{Dug}, and basic combinatorial topology \citep[e.g.][]{ST}.

\vspace{-4mm}
\subsection{On efficiency of Convexity Checkers}
\vspace{-4mm}
 Suppose all computations are done with floating point
arithmetic. More formally, consider the random access machine (RAM) with unit cost model of computation, where all four arithemtic operations are included in the instruction set. As usual,
$\ssf_{i\:j}$ denotes the number of incidences between $i$-dimensional and
$j$-dimensional faces and $\ssf_i=\ssf_{i\:i}$ stands for the number of $i$-dimensional
faces. Consider the case of $\R^3$. If  condition (CO) holds,  the algorithms by \citet{M96b} and \citet{Dev}, have the same time-complexity of $O(\ssf_{0})$. Our algorithm also has the complexity of  $O(\ssf_{0})$,  without requiring the assumption (CO); furthermore, its working  does not depend on whether the surface is orientable or not.
In a more general situation, where the conditions (CO) and (S), or some of them,
cannot be assumed, the complexity of our algorithm, as well as the algorithms of the
previous authors heavily depend on the following three factors: i) the combinatorics of
the cell-partition (e.g. simplicial or not), ii) the geometry of the realization (e.g.
generic positions of the vertices vs. completely general case), iii) the form of the
input.

Regarding iii), for example, the combinatorial  information about the input can be
given in the form of the complete poset of faces, or some subposet of faces, such as,
e.g. the vertex-facet graph. Furthermore, certain additional topological information
(e.g. the knowledge that the hypersurface is orientable, or a cyclic order of the facets at each
$(n-3)$-face) might speed up the convexity verification. The geometry of the
realization can be given by the equations of the facets, or by "coherent" inequalities
for the facets (CO), or by positions of the vertices, or in the form of inner normals
at $(n-3)$-cells to the $(n-2)$- and $(n-3)$-cells.

Now let us consider the general problem of convexity verification.  At one end is the
simplified setup, where the input hypersurface is simplicial and the realization is
sufficiently generic so that floating point arithmetic can safely be used. Under these assumptions
everything is fast, regardless of  the method used. On the other end is the completely
general setup, where nothing can be assumed.  One can also consider ``intermediate" models,
such as, e.g., where the input hypersurface is simplicial, but the positions of the vertices
are not necessarily generic. Another reasonable assumption would be that the hypersurface is not necessarily simplicial, but the realizations of the facets are known to be convex. We give our algorithm for the
most general case, where nothing can be assumed.  One of the motivations for this
generality is the work of Joswig and Ziegler (2004), who clearly demonstrated that from
the complexity theory point of view the convex hull problem is most interesting when we
cannot assume that the vertices (or hyperplanes) are in general position  -- or that  the
dimension of the space is a small fixed number.

If sufficient  linear-algebraic data and face incidence information are given about the
stars of $(n-3)$-faces (see Section \ref{sec:pseudocode} for details), then the
complexity of our algorithmic approach is still polynomial (in the Turing machine
model). The distinguishing  features of our approach are: 1) locality of testing and 2)
independence from the global topology of the input surface. Moreover, our approach
generalizes to piecewise-algebraic surfaces (this work is in progress).

\vspace{-4mm}
\section{Definitions and Background}\label{sec:definitions}
\vspace{-4mm}
From now on $\X^n$ is used to
denote $\R^n$ or $\SSS^n$. By a \emph{subspace} of $\X^n$ we mean an affine subspace in
the case of $\R^n$, and the intersection of $\SSS^n \subset \R^{n+1}$ with a linear
subspace of $\R^{n+1}$ in the case of $\SSS^n$. A \emph{hyperplane} is a subspace of
codimension one. A set $K \subset \X^n$ is called \emph{convex} if for any $x,y \in K$ there is
a minimal geodesic segment with end-points $x$ and $y$ that lies in $K$.  Then, the
dimension of $K$, $\dim K$, is understood as the dimension of a minimal subspace
containing $K$ (since such a subspace is unique for both $\R^n$ and $\SSS^n$, we denote
it by $\aff K$, even in the spherical case). We use $\inter K$ to denote the interior
of $K$ in $\aff K$: in other words, $\inter K$ stands for the \emph{interior} of $K$
\emph{ relative} to $\aff K$.  A \emph{hypersurface in } $\X^n$ is a pair $(\cM,r)$
where $\cM$ is a manifold of dimension $n-1$, with or without boundary, and $r:\cM \rightarrow \X^n$ is a
continuous \emph{realization} map. Unless specified otherwise, a \emph{manifold} will always mean a manifold without boundary. When the realization map is fixed we do not mention
it every time. A realization $r$ is called complete if any  sequence on $\cM$, which is
Cauchy with respect to the $r$-metric on $\cM$ (see Section \ref{sec:PL-immersions}),
converges to a point of $\cM$. A map  $i:\cM \rightarrow \X^n$ is called an \emph{
immersion} if $i$ is a local homeomorphism, in such a case we may also refer to
$(\cM,i)$ as an immersed hypersurface. A map  $f:\cM \rightarrow \X^n$ is called an \emph{embedding} if $f$ is a homeomorphism onto $f(\cM)$.  Obviously, an embedding is an immersion, but not vice versa. We call a
$k$-submanifold (with or without boundary) $\cS$ of $\cM$ flat with respect to the
realization map $r$   if $r:\cS \rightarrow r(\cS)$ is an embedding into a $k$-subspace
of $\X^n$.  A submanifold is called open if it is non-compact and without boundary.


 A \emph{convex body} in $\X^n$ is a closed convex set of full dimension. The hypersurface $(\cM,r)$ is called \emph{locally convex at} $p \in \cM$ if $p$
has a neighborhood  $\cN_p \subset \cM$ such that $r:\cN_p \rightarrow r(\cN_p)$ is a homeomorphism and $r(\cN_p)$ lies on the boundary of a convex body $K_{p}=K_{p}(\cN_p)$.
 Often, when it is clear from the
context  that we are discussing the properties of $r$ near $\mathbf{p}=r(p)$, we  say that $r$ is
convex at $\mathbf{p}$. If $K_{p}$ can be chosen so that $K_{p} \setminus r(p)$ lies
in an open half-space defined by some hyperplane  passing through  $r(p)$,  the
realization $r$ is called \emph{strictly convex} at $p$. We will also sometimes refer
to $(\cM,r)$ as strictly convex at  $r(p)$. By a theorem of \citet{Bu}
$\cN_p$ and $K_{p}(\cN_p)$  can always be chosen to allow a support hyperplane $H$ at $r(p)$ such
that  the orthogonal projection of $r(\cN_p)$ onto $H$ is an open $(n-1)$-ball.  Due to the local nature of this theorem, it holds in all spaces of constant
curvature.  When $K_{p}$ and $\cN_p$ satisfy this assumption, we refer to $K_{p}$ as a
\emph{convex witness} for $p$.

  Let us recall \citep[e.g.][]{Rock} that a point $p$
on the boundary of a convex set $K$ is called \emph{exposed } if $K$ has a support
hyperplane that intersects $\overline{K}$,  the closure of $K$, only at $p$; also, $p$ is
called \emph{extreme } if it does not belong to the interior of any interval contained
in $\partial K$. Thus, an \emph{exposed} point on a convex body $B$ is a \emph{point of
strict convexity} on the hypersurface $\partial B$. Conversly, a point of strict convexity
$p\in \cM$ for $(\cM,r)$ is an exposed point for a convex witness $K_p$. Local
convexity can be defined in many other, non-equivalent, ways (e.g., see van Heijenoort,
1952).

We will say that the hypersurface $(\cM,r)$ is \emph{the boundary of a convex body} $K \subset \X^n$ if $r$
is a homeomorphism from  $\cM$ onto $\partial K$. Hence, we exclude the cases when
$r(\cM)$ coincides with the boundary of a convex body but $r$ is not injective. Our
algorithm for PL-hypersurfaces will always detect a violation of the immersion property; in the  case
where $r(\cM)$ is the boundary of a convex body, but $r$  is not a homeomorphism, it will
produce the negative answer without trying to determine if $r(\cM)$ is the boundary of
a convex body. Of course, the algorithmic and topological aspects of this case may be
interesting to certain areas of  geometry, such as origami. Note that for $n\ge
3$ a closed $(n-1)$-manifold $\cM$ cannot be immersed into $\R^n$ by a non-injective
map $r$ so that $r(\cM)$ is the boundary of a convex set -- any convex hypersurface
in $\R^n$ is simply-connected and any covering map onto a simply-connected manifold
must be a homeomorphism. However, such immersions cannot be easily ruled out in the hyperbolic space $\HH^n$, as there are infinitely many topological types of convex
hypersurfaces in $\HH^n$ for $n > 2$ (Kuzminykh, 2005).

This paper is mainly focused on convexity of piecewise-linear (PL) hypersurfaces, in
particular, boundaries of polytopes. Denote by $B^d$ the closed unit ball at the origin
in  $\R^d$. A (disjoint) countable partition $\mathcal{P}$ of a topological space $\cM$
is called a \emph{semiregular cell-partition}  if (1) each element $C \in \mathcal{P}$,
called a cell of $(\cM,\mathcal{P})$, is homeomorphic to $\inter B^{\dim C}$, where
$\dim C \in \mathbb{N}$ and $\dim C \le \dim \cM$; (2) the closure $\overline{C}$ (in
$\cM$) of each $C \in \mathcal{P}$ is the union of $C$ and cells of smaller dimensions;
(3) for each $C \in \mathcal{P}$ there is a mapping $r_C:\overline{C} \rightarrow
B^{\dim C}$ which is a homeomorphism onto $r_C(\overline{C})$ and whose restriction  to
$C$  is a homeomorphism onto $\inter B^{\dim C}$. Authors that prefer to deal with
closed cells refer to cell-partitions as \emph{cell-complexes.}

If each cell is contained in the closure of finitely many cells, the partition is
called \emph{star-finite.} $\Star F$ denotes the subcomplex that consist of all (relatively
open) cells whose closure contains $F$.  If the closure of each cell  is the union of
finitely many cells, the partition is called \emph{closure-finite.} When a partition is both
closure- and star- finite, it is called \emph{locally-finite.}  Often, in the definition of
cell-partition (cell-complex) one insists on that each closed cell is the image of a
closed ball, which forces the compactness for each closed cell -- we do not make such a
requirement. Hence, in our definition the closures of the cells can be ``semiclosed-semiopen". Our notion of semiregular cell-partition is a natural generalization of the standard notion of regular cell-partition, also known as regular CW-complex, introduced  by J.H.C. Whitehead  \citep[see e.g.][]{Z}. Namely, a regular cell-partition is a semiregular locally-finite cell-partition, where the closure  of each cell is homeomorphic to a closed ball.

According to our definition, for example, the vertical projection on the plane of the
graph (in $\R^3$) of a continuous piecewise-linear (more formally, piecewise-affine)
function $f$ on $\R^2$, which is defined by finitely many affine equations and
inequalities, naturally induces a semiregular cell-partition $\mathcal{P}_f$ of $\R^2$:
each 2-cell is a maximal linearity set of the function. However,   $\mathcal{P}_f$
fails to be a regular cell-partition due to its unbounded cells. Semiregular partitions
are especially well-suited for the study of topology of real semialgebraic and
subanalytic sets -- any such set has a canonical finite semiregular cell-partition.

It is easy to see that any regular cell-partition can be subdivided into a triangulation. Thus, a manifold
admitting a regular cell-partition belongs to the category of PL-manifolds. It can be proven
that a semi-regular locally-finite cell-partition can also be subdivided into a triangulation. However,
 since the algorithmic part of this paper deals only with finite partitions, we omit
this theorem.

With any semiregular cell-partition there is a natural structure of poset. Namely, for cells $F$
and $C$ we have $F \preceq C$ if and only if  $F \subset \overline{C}$; we write $F \prec C$ if $F \subset \overline{C}$ but
$F \neq C$. If
$F \preceq C$, we say that $F$ is a face of $C$. We will use the same symbol $\mathcal{P}$ for a
partition and its poset. It is convenient to augment the
poset $\mathcal{P}$ with an infinum $\bot$, which is $\emptyset$, and a supremum
$\top$, which can be thought of as all of $\cM$. We use $\mathrm{Sk}_d(\cM,\cP)$ to
denote the $d$-skeleton of $(\cM,\cP)$  - i.e.  the subcomplex of  $(\cM,\cP)$ that
consists of all cell of dimensions not exceeding $d$.

 A subset of $\R^n$ is called polyhedral if it is defined by
a propositional formula in the language of the reals ($\R$) that uses only affine equations and
inequalities. A subset $S$ of $\SSS^n \subset \R^{n+1}$ is called polyhedral if $S= \SSS^n \cap
E$, where $E \subset \R^{n+1}$ is  defined by a propositional formula in the language of the
reals ($\R$)  that uses only linear (homogeneous) equations and inequalities.
 A \emph{PL-hypersurface} in $\mathbb{X}^n$ is a triple $(\cM,\mathcal{P};r)$,
where $\cM$ is a manifold with a semiregular cell-partition $\mathcal{P}$, and $r : \cM
\rightarrow \mathbb{X}^n$ is a continuous \emph{realization} map, such that  for  each
$C \in \mathcal{P}$   the set $r(\overline{C})$
 is polyhedral and is homeomorphic to $\overline{C}$.
 Note that although $r$ need
not even be an immersion, the restriction of $r$ to the closure of any cell $C$ of
$\cM$ must be an embedding.  The realization is called cellwise-flat  if the dimension
of the affine span of $r(\overline{C})$ is equal to $\dim C$. Although the term
\emph{face} is used  both for abstract cells and their geometric realizations, we
usually apply it for the realizations. If $\dim C=k$,  then $r(C)$ is called a $k$-face
of $(\cM,\mathcal{P};r)$.  Throughout the paper all faces, just as all cells in
topological partitions, are assumed to be \emph{relatively open.} At times we refer to
$(n-1)$-faces as \emph{facets,} $(n-2)$-faces as \emph{ridges,} and $(n-3)$-faces as
\emph{corners.} We may also use these geometric names for the underlying cells of
$(\cM,\mathcal{P})$ -- the meaning will always be clear from the context.

Let us consider  a connected \emph{PL-hypersurface}  $(\cM,\mathcal{P};r)$, where $r:\cM \rightarrow \mathbb{R}^n$ maps each cell onto a set of the same affine dimension, i.e., $\dim C = \dim \aff r(C)$. Suppose
$r:\cM \rightarrow \mathbb{R}^n$  has at least one point of strict convexity;
also, suppose that $r$ is locally convex at all points of all corners of $(\cM,\mathcal{P})$. Notice that if the last condition holds
for some point of a corner, it holds for all points of the corner. Our main Theorem \ref{theorem:main} states  that under these conditions $r$ is a homeomorphism on the boundary of a convex body.  This theorem
implies a test for global convexity of a PL-hypersurface that proceeds by checking the local convexity on each
of the corners.   The pseudo-code for the algorithm is given in Section
\ref{sec:pseudocode}. The complexity of this test depends not only on the model of
computation, but also on the way the surface is given as input data.  Let the input be
the poset of faces of dimensions $n-1$ (facets) $n-2$ (ridges) and $n-3$ (corners).
Suppose for each corner-ridge incidence $(C,R)$ we are given a Euclidean inner normal
to $r(R)$ at $r(C)$, and for each corner-facet incidence $(C,F)$ we are given a
Euclidean inner normal to $r(F)$ at $r(C)$. If we adopt the algebraic complexity  model
where each of scalar operations  \{comparison, addition, subtraction, multiplication, and
division\} has unit cost,  the complexity of the algorithm is
$O(n\ssf_{n-3\:n-2})=O(n\ssf_{n-3\:n-1})$. Complexity under other models is studied in
Section \ref{sec:Complexity}; we also study the complexity of extracting the
required input information from more common input representations.

The algorithm consists of $\ssf_{n-3}$ independent subroutines corresponding to the
$(n-3)$-faces, each with complexity not exceeding big-$O$ in the number of $(n-2)$-cells
incident to the $(n-3)$-face. In addition to the algorithmic implications, our generalization implies
that any $(n-3)$-simple compact PL-hypersurface in $\mathbb{R}^n$  is the boundary of a convex
polytope.

\vspace{-2mm}
\section{Geometry of Locally-Convex
Immersions}\label{sec:PL-immersions} \vspace{-2mm}

Recall that a path joining points $x$ and $y$ in a topological space ${\cal T}$ is a
continuous map  $p:[0,1] \rightarrow {\cal T}$. Such a path is called an \emph{arc} if
$p$ is injective. Denote by $\Arcs_{\cM}(x,y)$ the set of all arcs joining $x,y \in
\cM$.

An immersion $i: \cM \rightarrow \X^n$ induces a metric $d_{i}$ on $\cM$ by

\[ d_i(x,y)=\underset{a \in {\Arcs_{\cM}(x,y)}}{\inf} { |i(a)| } \]

\par \noindent where $|i(a)| \in \R \cup \infty$ stands  for the length of the
$i$-image of an arc $a$ joining $x$ and $y$ on $\cM$. This metric is called the
$i$-metric. Of course, for a general continuous realization $r$ it is not clear \emph{a priori} that there is a path  of finite length on $r(\cM)$ joining $r(x)$ and $r(y)$.
That is why we need Lemma \ref{lemma:arcwise}.

\begin{lemma}\label{lemma:arcwise} If $i$ is an immersion, then any two points of $\cM$ can be connected by an arc of a finite length. In particular,  $\cM$  is not only connected, but also arcwise connected.
\end{lemma}

The following lemma is used implicitely throughout our proofs and is also important for
understanding Van Heijenoort and Jonker \& Norman' arguments that we are employing.

\begin{lemma} If\  $i$ is an immersion, then the metric topology defined by the $i$-metric is equivalent to the original topology on $\cM$.
\end{lemma}

\par \noindent Van Heijenoort's (1952) proofs of these two lemmas, given for $\X^n=\R^n$, work
for $\X^n=\SSS^n$ without changes. Note that since $i$ is an immersion, then for a
``sufficiently small"  subset $\cS$ of $\cM$ the map $i \vert_{\cS}$ is a
homeomorphism and, therefore, the topology on $\cS$ that is induced by the metric topology of $\X^n$ , is equivalent to the
intrinsic topology of $\cS$, i.e., the subspace topology. Thus, for sufficiently small
subsets of $\cM$ the three topologies considered in this section  are
equivalent. This fact will be used on numerous occasions without an explicit reference
to the above lemmas. We will also need the following theorem.

\begin{theorem} \emph{(Van Heijenoort, 1952)}\label{theorem:H}
If a complete locally convex immersion $\ff$ of a connected manifold $\cM$
($\dim~\cM=n-1$) into $\R^n$ ($n\ge3$) has a point of strict convexity, then $\ff$ is a
homeomorphism onto the boundary of a convex body.
\end{theorem}

 Recall that a
subset $C$ of  the linear space $\R^{n+1}$ is called a \emph{cone} if
$\lambda C \subset C$  for any $\lambda \in \R_{+}$. A cone $C$ is called pointed if $\mathbf{0} \subset C$. A
cone is called salient if it does not contain any linear subspace except for
$\mathbf{0}$. If $S \subset \R^{n+1}$, then we denote by $p \cdot S$ the cone with apex
$p$ over $S$.

Let $x$ be a point on $\cM$ and let $S$ be a subset of $\X^n$ such that $x \cap r^{-1}(S)\neq \emptyset$ for a realization map $r:\cM \rightarrow \X^n$. We denote by
$r^{-1}_xS$ the connected component of $r^{-1}(S)$ that contains $x$. Consider now the case of $\X^n=\SSS^n$. We assume
that $\SSS^n$ is embedded as the standard unit sphere into $\R^{n+1}$.
For a point $a \in \SSS^n \subset \R^{n+1}$  we denote by $c_{a}$ the central projection mapping from the half of $\SSS^n$ that is centered at $a$ onto the tangent plane $\mathbf{T}_{a} \subset \R^{n+1}$ to  $\SSS^n$ at $a$. To simplify the visual appearence of formulas we will use $c_x$ instead of $c_{r(x)}$ in  a context where the map $r:\cM \rightarrow \SSS^n$ is fixed.

The following theorem, whose proof follows the approach taken by Jonker \& Norman (1972),  shows that for spherical immersions absence of a point of a strict
convexity cannot result in the loss of  global convexity, as it happens in the Euclidean case.
\begin{theorem}\label{theorem:strict_or_conical}
Let $i: \cM \rightarrow \SSS^n$ ($n\ge3$) be a locally convex complete immersion of a
connected $(n-1)$-manifold $\cM$.  Then $(\cM,i)$ is strictly locally convex in
at least one point, or $i(\cM)= \partial(\mathbb{S}^n \cap C)$, where $C$ is a pointed
convex cone in $\mathbb{R}^{n+1}$.
\end{theorem}
In the latter case we say that $i$ is \emph{conical.}

\begin{proof}  If $\cM$ has a point $p$ such that  $i(p)$  is an extreme point for some convex witness $K_{p}$,
then, by a classical theorem of Straszewicz \citep[e.g.][]{Rock}, either $i(p)$  is also an
exposed point, or there are infinitely many exposed points of $K_{p}$ arbitrarily
close to $i(p)$. Since an exposed point is a point of strict convexity, the surface $(\cM,i)$ is
strictly  convex in at least one point and the theorem follows.

Otherwise, suppose that for all $p \in \cM$  the image $i(p)$ is an interior point of a segment on $i(\cM)$. Below we will show
that this assumption implies that $i$ is a homeomorphism onto $\partial(\mathbb{S}^n \cap C)$
for some pointed convex cone $C$.
 If a segment through $i(p)$ cannot be extended to a circle on $i(\cM)$, which is the $i$-image of a closed curve through $p$ on
 $\cM$, then pick an end point of a maximal segment $I_p$ through $i(p)$ and call it $i(a)$. Since $i(a)$ is not extreme, it must lie in the
 interior of another segment. There is a support plane $H_a$ for $a$  such that $\inter I_p$  and $H_a$ do not intersect near $i(a)$. Consider now $i_a^{-1}H_a$, which is
  either  a closed submanifold of dimension at least 1 or a closed submanifold with boundary.
 If $i_a^{-1}H_a$
 is a submanifold with boundary, then the central projection $c_a i (i_a^{-1}H_a)$ of $i (i_a^{-1}H_a)$ on $\mathbf{T}_a \cong \R^{n}$ must have parallel lines on the boundary, which correspond to
intersecting
 half-circles on $\SSS^n$. Furthermore, through any two points on the relative boundary of $c_a i (i_a^{-1}H_a)$ there are
two parallel lines contained in the relative boundary of $c_a i (i_a^{-1}H_a)$.  Let $R
\subset \mathbf{T}_a$ be a maximal subspace contained in $c_a  i( i_a^{-1}H_a)$.
 The $c_a$-preimages of points at infinity of $R$ form a flat closed submanifold of  $\cM$.  Thus, $\cM$ must have a flat closed submanifold of
 dimension at least 1.

 Let $\cF \subset \cM$ be a closed flat submanifold of maximal dimension. To simplify the visual
appearance of formulas, denote by $F$ the $i$-image of $\cF$, and by $F^{\perp}_y$ the
orthogonal complementary subspace to $F$ in $\SSS^n$ at the point $r(y)$. If $\dim \cF=n-1$, then $\cF=\cM$ and the proof is completed.

Let us turn now to the case of $\dim \cF<n-1$.  Since $(\cM,i)$ is locally convex and
 does not have points of strict convexity, there is an open set $\cU_{\cF} \subset \cM$ that contains $\cF$ such that $i(\cU_{\cF})$
 is of the form $\partial (\mathbf{0} \cdot F \times C_F) \cap \SSS^n$, where $C_F$
 is a salient $(n-1-\dim \cF)$-dimensional convex cone in $\mathbf{0} \cdot F^{\perp}_y$.  Note that
 $i_y^{-1}(F^{\perp}_y \cap i(\cM))$ is a locally convex hypersurface in $F^{\perp}$, which is
strictly
 convex at $y$.  If $\dim F^{\perp}=2$, then $i$ is conical -- the pointed cone $C$ is the product of a linear  $(n-1)$-subspace in $\R^{n+1}$   and two rays (with origin at $\mathbf{0}$) in a complimentary linear 2-subspace.

Suppose now  $\dim F^{\perp}_y>2$. Upon applying van  Heijenoort's theorem to
the map \linebreak $c_y  i: i_y^{-1}(F^{\perp}_y \cap i(\cM)) \rightarrow c_y(F^{\perp})$,
we see that $c_y  i\left[i_y^{-1}(F^{\perp} \cap i(\cM))\right]$ is a complete convex hypersurface
in $c_y(F^{\perp}_y) \subset \mathbf{T}_y$. But this is true for all $y \in \cF$. Furthermore,
if $y, y' \in \cF$ are sufficiently close, the local convexity implies that
$M_y=i\left[i_y^{-1}(F^{\perp}_y \cap i(\cM))\right]$ and $M_{y'}=i\left[i_{y'}^{-1}(F^{\perp}_{y'} \cap i(\cM))\right]$ are
isometric: an isometry can be chosen as a
minimal rotation $\rho_{yy'}$ of $\SSS^n$ in $\R^{n+1}$ around the orthogonal
complement  to the plane spanned by vectors $r(y)$ and $r(y')$, which  maps $r(y)$ to $r(y')$. Since $\cF$ is compact, the manifolds $M_y$ are isometric
for all values of $y \in \cF$. Also, for any   $y, y' \in \cF$ the isometry mapping $\rho_{yy'}$ from $M_y$ to $M_{y'}$ preserves the distance of each
point of $M_y$ to $F$.
If  the map $i:i_y^{-1}(F^{\perp}_y \cap i(\cM)) \rightarrow F_y^{\perp}$ is strictly convex at
some point $z \neq y$, then any geodesic segment $I_z=i(\cI_z)$, where $\cI_z$ is a flat open
1-submanifold  through $z \in \cM$, is transversal to $F^{\perp}_y$. Therefore, any support
plane $H_{I_z}$ for $(\cM,i)$ through $I_z$ intersects with a support plane for $(\cM,i)$ through $F$ over
a proper subset  of $F$. But then for any neighborhood $\cU_y$ of $y$ there will
be $y' \in \cU_y \cap \cF$ such that $\rho_{yy'}$ will move $r(z)$ to a point lying on the
other side of $H_{I_z}$ with respect to  the convex witness of $z$, resulting in a contradiction. Thus, for all $y \in \cF$
the hypersurface $c_y  i\left[i_y^{-1}(F^{\perp} \cap i(\cM)) \subset c_y(F^{\perp}_y)\right]$
cannot be strictly convex at any $z\neq y$, which means  $c_y  i\left[i_y^{-1}(F^{\perp} \cap i(\cM))\right]$ is a convex salient cone
in $c_y(F^{\perp}_y)$. This implies that $i$ is conical.
\end{proof}

The notion of convex part, introduced by Van Heijenoort (1952), happens to be very
useful in working with local convexity. A \emph{convex part} of  $(\cM,i)$ at a point
of strict convexity $i(o)$ (i.e. containing a point of strict convexity $i(o)$), is a
connected subset $C$ of  $i(\cM)$, with $i^{-1}_oC$ open in $\cM$, such that
\par (1) $i(\partial i^{-1}_oC)= H \cap i(\cM)$, where $H$ is some hyperplane in $\X^n$, not passing through $i(o)$,
\par (2) $C$ lies on the boundary of a convex body $K_C$ such that $\partial K_C \subset C \cup H$.

The set $i^{-1}_oC$ is called an \emph{abstract convex
part at} $o$  and is denoted by ${\mathcal C}$.
We will denote $i(\partial i^{-1}_oC)$ by $\rel \partial C$ and we will call $H \cap K_C$ the
\emph{lid} of the convex part $C$. Let $H_o$ be a supporting hyperplane at $i(o)$. Let us call
the \emph{open} half-space defined by $H_o$ that contains $C \setminus i(o)$ the \emph{positive
half-space} and denote it by $H^+_o$.

\begin{theorem} (after van Heijenoort)\label{theorem:existence_of_convex_part}
If an immersion of an $(n-1)$-manifold  ($n\ge3$) into $\X^n$ has a point of strict
convexity, it has a  convex part containing this point.
\end{theorem}
\begin{proof}van Heijenoort's proof  works for $\X^n$, $n\ge3$, without changes.
\end{proof}

Let us fix a supporting hyperplane $H_o$ and consider a family of hyperplanes  such
that: (1) they are pairwise disjoint in $H^+_0$ and their intersections with $H^+_0$
form a  partition of $H^+_0$, (2) they are  all orthogonal to a line (circle) $L$,
transversal to $H_o$, (3) they do not contain $i(o)$. Let us call such a family the
fiber bundle $\{H\}_{(L,H_o)}$ of the positive half-space defined by $L$ and $H_o$.
Denote by $\lambda > 0$  the length of the segment on \emph{the
line (circle) $L$} from $l = L \cap H_o$ to $H \in \{H\}_{(L,H_o)}$, called the height of $H$. The
elements  of $\{H\}_{(L,H_o)}$ can then be indexed by their heights. We use
$H_{\lambda}$ for the hyperplane of height $\lambda$. For $\X^n=\R^n$ the range of
$\lambda $ is $(0,+\infty)$ and for $\X^n=\SSS^n$ it is  $(0,\pi)$.
If there exists  a convex part  at $i(o)$, then  $\{H_{\lambda}\}_{(L,H_o)}$  defines a
family $\{C_{\lambda}\}$ of convex parts at $i(o)$, ordered by inclusion.  Each such convex part is
"squeezed" between $H_o$ and $H_{\lambda}$ and inherits the height from its
$H_{\lambda}$. Let us now consider the union $\text{\bf{\itshape C}}_o$ of all convex parts at $i(o)$: \emph{ we
want to prove that $\text{\bf{\itshape C}}_o$ is a convex part itself.} This statement was proven
by van Heijenoort for $\X^n=\R^n$ ($n\ge 2$) and we will prove it for $\SSS^n$ ($n\ge 3$).

\begin{theorem}\label{lemma:convex_alternative}
 Let $i: \cM \rightarrow \SSS^n$ be as in Theorem
 \ref{theorem:strict_or_conical}, and  let $C=i(\cC)$ be
 a convex part  at  $i(o)$, defined by a hyperplane
 $H_{\lambda} \subset \SSS^n$ ($\lambda \neq 0$) from a fiber bundle
 $\{H_{\lambda}\}_{(L,H_o)}$, where $L \perp H_o$.
 Suppose  $\rel \partial C$ is the boundary of a convex
 set $S$ in $H_{\lambda}$. Then  either $S$ is the
 $i$-image of a topological disk ${\cal S}$ ($\dim {\cal S} \le n-1$) in
 $\cM$ and $\cM={\cal C} \cup {\cal S}$,  or $C$
 is a proper subset of a larger convex part  at
 $i(o)$ and defined by the same bundle
 $\{H_{\lambda}\}_{(L,H_0)}$.
\end{theorem}

\begin{proof} We will now prove this theorem by a perturbation argument, which reduces the spherical
case to that of $\R^n$. Since $\rel \partial S= \rel \partial C$, $S \subset
H_{\lambda}$ $(\lambda>0)$, and $\dim S=n-1$, we conclude that $\overline {\conv C}
\cap H_o=i(o)$. Since, by Lemma \ref{lemma:arcwise}, $\cM$ is arcwise connected, all of
$\overline{\conv C}$, except for $i(o)$, lies in the positive halfspace $H_o^+$.  Thus,
there is a hyperplane $H$ in $\SSS^n$, arbitrarily close to $H_o$ and orthogonal to
$L$, such that $C$ lies in an open halfspace $H^+$ defined by $H$. Let $p$ be the pole
of $\SSS^n$ with respect to $H$ that lies in $H_o^+$, and let $\fc_{p}:H^+ \rightarrow
\mathbf{T}_{p}$ be the central projection map on the tangent plane $\mathbf{T}_{p} \subset \R^{n+1}$.
The set $i^{-1}_o (H^{+} \cap \cM )$  is obviously a manifold. The map $\fc_{p}  i$ is a
locally-convex immersion of $i^{-1}_o (H^{+} \cap \cM )$ into $\mathbf{T}_{p}$ with a point of
strict convexity, $i(o)$.  Any Cauchy sequence on $\cM$ under the $\fc_{p} i$-metric is
also a Cauchy sequence under the $i$-metric. Thus the immersion map $\fc_{p}  i:i^{-1}_o (H^{+} \cap \cM
) \rightarrow \mathbf{T}_{p}$ is complete and, therefore, satisfies the conditions of Theorem
\ref{theorem:existence_of_convex_part}.  It maps any (spherical) convex part
centered at  $i(o)$ onto a Euclidean convex part; it also maps the fiber bundle
$\{H_{\lambda}\}_{(L,H_o)}$ to a  fiber bundle in $\mathbf{T}_p$. In the case of $\X^n=\R^n$ the
statement of the theorem is known (van Heijenoort). We conclude that either   $\cM={\cal C} \cup
{\cal S}$, or $C$ is a proper subset of a larger convex part centered at $i(o)$ and
defined by the same bundle $\{H_{\lambda}\}_{(L,H_0)}$. \end{proof}

\vspace{-2mm}
\section{From Local to Global Convexity on the Sphere}
\vspace{-2mm}
Let us recall that Van Heijenoort proved that a complete locally convex
immersion $\ff$ of a connected manifold $\cM$ ($\dim~\cM=n-1$) into $\R^n$ ($n\ge3$) is
the boundary of a convex body, if $\ff$ has a point of strict convexity. For $n=3$ this
result, according to van Heijenoort, follows from four theorems in Alexandrov's book
(1948). Jonker \& Norman (1973) proved that if $\ff$ does not have a point of strict
convexity, $\ff(\cM)$ is the direct affine product of a plane locally convex curve and
a subspace $L \cong \R^{n-2}$ of $\R^n$.

\begin{theorem}\label{theorem:sphere}
Let $i : \cM \rightarrow \SSS^n$ $(n\ge2)$ be an immersion of a connected
$(n-1)$-manifold $\cM$, satisfying the following conditions:
\noindent  1) $\cM$ is complete with respect to the $i$-metric,
\noindent  2)  $(\cM,i)$ is locally convex at all points of $\cM$.
Then $i : \cM \rightarrow \SSS^n$ is an embedding onto the boundary of a convex body in $\SSS^n$.
\end{theorem} \begin{proof}
If points of strict convexity are absent, then by Theorem \ref{theorem:strict_or_conical} the map  $i : \cM \rightarrow \SSS^n$ is an embedding onto the boundary of a convex body. Thus,  we can assume that a point of strict convexity exists. If $o \in \cM$ is such a point,
then by Theorem \ref{theorem:existence_of_convex_part} there is a convex part
containing $i(o)$. Consider the union $\text{\bf{\itshape C}}_o$ of all convex parts at
$i(o)$. Denote by $\zeta$  the least upper bound of the  heights  of convex parts
defined by $\{H_{\lambda}\}_{(L,H_o)}$   at $i(o)$. By Theorem \ref{lemma:convex_alternative}  $\partial \text{\bf{\itshape C}}_o
\subset H_{\zeta} \cap i (\cM)$ and $\rel \partial \text{\bf{\itshape C}}_o$ is the
boundary of a closed convex set $D$ in $H_{\zeta}$. Two mutually excluding cases are
possible.
\par \noindent \texttt{Case 1:} $\dim D<n-1$. Then, following the argument of van
Heljenoort (Part 2: pp. 239-230, Part 5: p. 241, and  Part 3: II on p. 231), we
conclude that $\text{\bf{\itshape C}}_o \cup D$ is the homeomorphic $i$-image of the
$(n-1)$-sphere $i^{-1}(\text{\bf{\itshape C}}_o) \cup i^{-1} (D) \subset \cM$, where
$i^{-1}(\text{\bf{\itshape C}}_o)$ is a maximal abstract convex part at $o$. Since
$\cM$ is connected, $i^{-1}(\text{\bf{\itshape C}}_o) \cup  i^{-1} (D) = \cM$, and $i :
\cM \rightarrow \SSS^n$ is a convex embedding of $\cM$.

\par \noindent  \texttt{Case 2:} $\dim D=n-1$. By Theorem \ref{lemma:convex_alternative}
$\text{\bf{\itshape C}}_o$  is either a proper subset of a larger convex part, or
$\text{\bf{\itshape C}}_o$, together with the lid $D$, is the homeomorphic  $i$-image
of $\cM$. Since the former alternative is excluded by definition of $\text{\bf{\itshape
C}}_o$, $i : \cM \rightarrow \SSS^n$ is a convex embedding onto $\partial
(\text{\bf{\itshape C}} \cup D)$.
\end{proof}

Note that the statement of the above theworem is invalid for $n=2$. For example, although the 1-surface in $\SSS^2$ depicted in Figure 1 is
locally convex at all points, it does not bound any convex set on $\SSS^2$.
\begin{figure}[h]
\begin{center}
\resizebox{!}{160pt}{\includegraphics[clip=false,keepaspectratio=false]{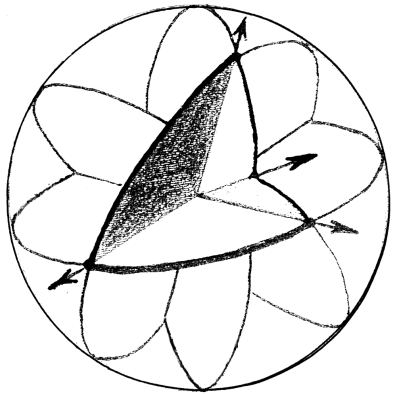}}
\caption{Locally-convex PL-hypersurface in $\SSS^2$, which is not convex}
\end{center}
\end{figure}
\vspace{-4mm}
\section{Locally convex PL-surfaces}
\vspace{-4mm}

Let $\mathcal{P}$ be a fixed star-finite semi-regular cell-partition of $\cM$.  Recall that in our terminology a cell is  always homeomorphic to an open ball. We say that $r$ is locally convex at a cell $C \in \cP$ if it is locally convex at each point of $C$.

\begin{theorem}\label{theorem:main}
Let $r: \cM \rightarrow \R^n$ $(n>2)$ be a complete cellwise-flat PL-realization of a
connected manifold $\cM$ $(\dim \cM = n-1)$ such that
\par \noindent  1) $r$ is locally convex in at least one point of each $(n-3)$-cell.
\par \noindent 2) $r$ is bounded or strictly locally convex in at least one point of $\cM$.

Then $r: \cM \rightarrow \R^n$ is an embedding onto the boundary of a convex body defined by
(possibly infinitely many) affine inequalities.

\end{theorem}
\begin{proof} Upon invoking the definition of cellwise-flat PL-realization and
that of local convexity, we conclude that $r$ is an immersion. We know that $r: \cM
\rightarrow \R^n$ is locally convex at all $(n-3)$-cells.  Since $r$ is a cellwise-flat
PL-realization, $r$ is also locally-convex at all $d$-cells for $d>n-3$.  If $r$ is bounded, then $\conv r(\cM)$ is bounded and compact.
By Straszewicz's
 theorem \citep{Rock}, the set $\conv r(\cM)$ has an exposed point. As $r$ is complete, this exposed point must the $r$ image of some $x\in \cM$. Since $r$ is locally convex at $x$, it is also strictly locally convex  at $x$. If we prove
that $r$ is locally convex at all cells, by Theorem $\ref{theorem:H}$ the map
$r: \cM \rightarrow \R^n$ is a convex embedding. We proceed by reverse induction in
cell's dimension.  Suppose we have shown that $r: \cM \rightarrow \R^n$ is
locally convex at each $k$-cell, where $0<k \le n-3$ (and therefore at all cells of
higher dimensions). If $n-3=0$, the proof is finished. So, let $n \ge 4$ and let us consider a $(k-1)$-cell $F \in \mathcal{P}$.  Consider $r(\Star F) \cap \SSS_F$, where $\SSS_F$ is a
sufficiently small $(n-k)$-sphere lying in an affine subspace complementary to $\aff r (F)$ and
centered at some  point of $r(F)$. Also note that $\dim \SSS_F=n-k \ge2$.
The map $r: \cM \rightarrow
\R^n$ is locally convex at $F$ if and only if the hypersurface  ${\mathbb S}_F \cap
r(\Star F)$ in  ${\mathbb S}_F$ is convex. Since   $r: \cM \rightarrow \R^n$  is
locally convex at each $k$-cell, the surface ${\mathbb S}_F \cap \Star F$  is locally convex at
each vertex and therefore locally convex everywhere. The set $r^{-1}({\mathbb S}_F \cap r(\Star
F))$ is complete in the $r$-metric and thus, by Theorem \ref{theorem:sphere}, the surface ${\mathbb
S}_F \cap r(\Star F)$ is an embedded convex hypersurface in $\SSS_F$. Notice that the condition $\dim \SSS_F=n-k > 2$
is essential to the applicability of Theorem \ref{theorem:sphere} (see Figure 1 for a locally convex surface in $\SSS^2$ which is not a convex surface in $\SSS^2$).
So, $r: \cM
\rightarrow \R^n$  is locally convex at $F$.

To recap, the above induction argument shows that $r$ is locally convex at all vertices, and, therefore, at
all points. The metric induced
 by $r$ is
indeed complete. Upon applying Van Heijenoort's Theorem \ref{theorem:H} and Theorem
\ref{theorem:sphere}, we conclude that $r$ is an embedding and that $r(\cM)$ is the boundary of a convex body.
\end{proof}

\begin{corollary}
Let $r: \cM \rightarrow \R^n$ $(n>2)$ be a complete cellwise-flat PL-realization of a
connected manifold $\cM$ $(\dim \cM = n-1)$. Suppose  $r$ is bounded or is strictly
locally convex in at least one point. If $(\cM,\mathcal{P})$ is $(n-3)$-simple, i.e.
exactly three $(n-1)$-cells make contact at each $(n-3)$-cell, then $r(\cM)$ is the
boundary of a convex polyhedron.
\end{corollary}

\vspace{-4mm}
\section{Convexity Checker for PL-hypersurfaces}\label{sec:pseudocode}
\vspace{-4mm}
In this section we present a polynomial-time algorithm for checking the convexity of
any PL-realization $r: \cM \rightarrow \R^n$  ($n\ge3$) of a  semi-regular cell-partition
$\mathcal{P}$ of a connected compact  $(n-1)$-manifold $\cM$. The map $r$ under testing is assumed to
be cellwise-flat (see Section \ref{sec:definitions}), which implies that each cell $C \in
\mathcal{P}$  is homeomorphicly mapped by $r$   to an open subset of an affine subspace
of dimension $\dim C$. We do not
assume that the realization is an immersion: if it is not an immersion, the algorithm will detect this. We
do not make any generic position assumptions.

In describing the algorithm we assume that certain combinatorial and geometric
information is readily available. This input information is exactly what should be kept
by a convex hull computer if it is to use our verification procedure. Later we discuss
the complexity of extracting the necessary input information from PL-surface
descriptions given in some typical  formats.

 If any of
the subprocedures return 1 or ``false", the main procedure returns ``false" as the
final answer.  The idea of the algorithm is to check that the immersion and the local
convexity properties hold at each corner. For each corner $C$ this check
 is reduced, roughly speaking, to the verification of convexity of a certain cone $K(r,\Star C)$  in
$r(C)^{\perp}\equiv \R^3$, which is constructed  from the poset of $\Star C$ and the restriction of $r:\cM \rightarrow \R^n$ to $\Star C$.  Such a cone is not
unique -- for example, any non-singular affine transformation of $K(r,\Star C)$ is just
as good as $K(r,\Star C)$.   This reduction from the star of a corner to a cone in
$\R^3$ is done by the procedure \textsf{Reduce-to-3D}.

\vspace{-4mm}
\subsection{\protect\large Input Conventions}\label{sec:pseudocode}
\vspace{-4mm}
Let  $X$ be a subset of $ \R^n$. We use $\aff X$ to denote the affine subspace
spanned by $X$ and $\overrightarrow{\aff X}$ to denote the \emph{linear} subspace
$\{x-x'\;\;\vline \;\;x,x'\in \aff X\}$.
 For a (not necessarily convex) polytope $\Pi \subset \R^n$ with a face $F$ an
\emph{inner normal} at $F$ is any vector $\n$ in $\overrightarrow{\aff \Pi}$ such that
for any point $p$ on $F$ there is $\epsilon>0$ such that $p+\epsilon \n \in \inter
\Pi$. The normal $\n$ is called a Euclidean normal if $\n \perp \aff F$.

As a reminder, we refer to the $(n-3)$- and  $(n-2)$-cells of the partition as
\emph{corners} and \emph{ridges} respectively, and we refer to $(n-1)$-faces as
\emph{facets}. We also use these terms to refer to the realizations of these cells in
$\R^n$.  Mathematically, the input is given as follows: \pagebreak
\begin{enumerate}
  \item the subposet $\mathcal{P}[n-3,n-2,n-1] \subset  \mathcal{P}$ of corners,
ridges, and facets where it is known in advance which are which;
  \item  a Euclidean inner normal to  $R$ at $C$ for each  ridge-corner incidence $(R,C)$;
  \item  a Euclidean inner normal to $F$ at $C$ for each   facet-corner incidence $(F,C)$.
\end{enumerate}
 The data in (1) will be referred to
as \emph{combinatorial}, and that, described in (2) and (3),  as
\emph{linear-algebraic}.
 We assume that each vector in the linear-algebraic data "knows" the corresponding abstract cells in
$\mathcal{P}$ , and that each abstract cell in  $\mathcal{P}[n-3,n-2,n-1]$ ``knows" all
normal vectors related to it. The  input data-structure can be implemented as
a double-linked adjacency list, with appropriate attribute fields for dimensional and
linear-algebraic data. Namely, we can create an adjacency list for the directed (multi-) graph whose vertex set consists of
elements of $\mathcal{P}[n-3,n-2,n-1]$  and whose edge set consists of all ordered pairs
$(\textrm{C,C}')$ and $(\textrm{C',C})$, where $\textrm{C} \prec \textrm{C}'$ or $\textrm{C}'
\prec \textrm{C}$ in $\mathcal{P}[n-3,n-2,n-1]$.  When the input is available in
this form we say that the \emph{input is given in the standard form.}

In applications a PL-hypersurface is usually specified by a subposet   of the face
poset, which includes the vertices or the facets or both; it is normally equipped
either with the coordinates of vertices or with the equations (or unequalities) for the
facets.   Suppose now the input is given as the poset
$\mathcal{P}[0,n-3,n-2,n-1]$, equipped with the
coordinates of the vertices; in this case  we will say that the input is given in \emph{traditional} form. If the partition $\cP$ is a triangulation, then the linear-algebraic data  required for our algorithm (standard form)  can be
produced in linear time in $\ssf_{n-3\;n-2}$, which is also polynomial in the total bit size of the input.
More  generally, if the  face numbers of facets of $(\cM,\mathcal{P})$ are bounded by a
universal constant (in $\ssf_{n-3\;n-2}$), the linear-algebraic data for the standard form of the input can be computed  by using
$O(\ssf_{n-2\;n-3})$ field arithmetic operations.

\vspace{-4mm}
\subsection{Preprocessing}\label{sec:pseudocode}
\vspace{-4mm}
  By \emph{preprocessing} in the context of problems of verification of geometric properties we
  mean any computation that does not depend on the geometric realization (in our case $r$),
  but only on the topology or combinatorics of the object (in our case -- the pair $(\cM,\mathcal{P})$).

Since $\cM$
  is a manifold, the facets of $(\cM,\mathcal{P})$ making contact at a corner are ``glued'' to each other in
  a circular fashion. Same can be said about the ridges. The circular structure of the stars of $(n-3)$-cells implies that for each
  $(n-3)$-cell $C$ we have $\ssf_{n-2}(\Star C)=\ssf_{n-3}(\Star C)$. The last formula implies that
  for the whole $\mathcal{P}$ we have $\ssf_{n-3\:n-2}=\ssf_{n-3\:n-1}$. More properly, a topologist would  say that
  the ``links'' (defined via the 1-skeleton of the \emph{dual partition}, a well-known construction
  going back to H. Poincare: see Seifert \& Threlfall, 1980) of the corners are \emph{circles}.
These circles can be thought of as
polygons whose vertices correspond to the facets of $\mathcal{P}$  and edges to the
ridges of $\mathcal{P}$. To apply our algorithm we need to determine a cyclic order of ridges
around each corner. Such  an order is unique up to the choice of direction..  To
apply our algorithm for different realizations of the same cell-partition of
$\cM$ it is reasonable to maintain a cyclic order of ridges around each corner; this can be considered as preprocessing.

\vspace{-4mm}
\subsection{Algorithm}\label{sec:pseudocode}
\vspace{-4mm}

 The main procedure \textsf{Convexity-Checker} is given in Algorithm 1. \textsf{Convexity-Checker} works on a stack \textit{Corners}, in which we put all $(n-3)$-faces of $(\cM,\mathcal{P})$ prior to starting. Subroutines \textsf{Reduce-to-3D}  and \textsf{Check-if-Cone-Convex} are used by \textsf{Convexity-Checker}, and  subroutine \textsf{Is-Folded} is used by \textsf{Reduce-to-3D}.  Before describing the working of the algorithm, let us define a couple of auxiliary notions.

For any $m\ge 2$ a graph with vertex set $V=\{0,1,\ldots,m\}$ ($m\ge2$) and edge set $E=\{(01),\ldots,(0m)\}\cup \{(12),\ldots,(m-1\:m),(m,1)\}$ is called the $m$-wheel graph and is denoted by  $W_m$.   Vertices $RimV(W_m)=\{1,\ldots,n\}$ are called rim vertices;  edges $RimE(W_m)=\{(12),\ldots,(m-1\:m),(m,1)\}$ are called rim ridges. Edges $\{(01),\ldots,(0m)\}$ are called spokes.  Vertex $0$ is called the center of $W_m$. Let $p:V\rightarrow \R^3$ be a realization of the vertex set of $W_m$ in $\R^3$. If $m\ge3$, then we can assign to each  3-cycle $(0\:i\:i+1)$ (where $i+1$ is taken~$\mod m$) in $W_m$ a geometric simplex  in $\R^3$ with the vertices $p(0)$, $p(i)$, and $p(i+1)$. Therefore, the map $p:V\rightarrow \R^3$ produces  a simplicial surface with boundary. With a slight abuse of terminology we will say that  a realization $p:V\rightarrow \R^3$ of the wheel graph $W_m(V,E)$ (where $m\ge 3$) is convex if $p$ is injective and the resulting simplicial 2-surface, which we denote by $p[W_m]$, is convex at $p(0)$. While we may encounter Euclidean realizations of the vertex set of $W_2$ , we will not have a need to associate a surface in $\R^3$ with such realizations.

Once a corner $C$ is popped
from the stack, a pair $(W_m,\mathbf{n})=(W_m[C],\mathbf{n}[C])$ is created. This pair
consists of the wheel graph $W_m=W_m[C]$, which encodes the combinatorics of $\Star C$, and an array of vectors $\mathbf{n}[C]$, whose elements are the  inner normals to the $r$-realizations of ridges and facets of $\Star C$ at the face $r(C)$. Namely,   $m$ is the number of ridges meeting at $C$,
the center of $W_m$ corresponds to $C$, the rim vertices correspond to  the ridges of $\Star C$, the rim edges
correspond to the facets of  $\Star C$, and the ``spokes" correspond to the
corner-ridge incidences.
\medskip
\begin{algorithm}\label{alg:main} \caption{\textsf{Convexity-Checker}} \begin{algorithmic}
\While{ $\textit{Corners} \neq \emptyset$} \State  Pop an $(n-3)$-face $C$ from
$\textit{Corners}$ \State Create $(W_m,\mathbf{n})$ for  $\Star C$

\If{$\textsf{Reduce-to-3D}(W_m,\mathbf{n})=1$} \Return{\emph{false}}
\Comment{\texttt{not convex}}

\Else \If{$\textsf{Reduce-to-3D}(W_m,\mathbf{n})\neq 0$}

\State {$(W_m,\fp)\leftarrow \textsf{Reduce-to-3D}(W_m,\mathbf{n})$}
\If{\textsf{Check-if-Cone-Convex-in-3D}$(W_m,p)$=\textit{false}} \Return{\emph{false}}
\Comment{\texttt{not convex}} \EndIf \EndIf \EndIf \EndWhile

\Return{\emph{true}} \Comment{\texttt{yes, convex}}
\end{algorithmic} \end{algorithm} \medskip
\clearpage
\subsection{\protect \emph{\textsf{Reduce-to-3D}} and its Subroutines}
\vspace{-4mm}
 For a
corner $C$ of $(\cM,\mathcal{P})$ let $\{0,1,\ldots,m\}$ be the  vertices of the wheel graph
$W_m=W_m[C]$, with $0$ as the center of $W_m$ (which corresponds to the conrer $C$), and let $[1,\ldots,m]$ be a cyclically
ordered list of the rim vertices of $W_m$ (which correspond to the ridges of $\Star C$). The
input to \textsf{Reduce-to-3D} consists of:  the wheel graph $W_m$;  the Euclidean inner
normals  $\n_{1},...,\n_{m}$ for  the $r$-images of ridges of $\Star C$ at $r(C)$; the Euclidean inner
normals  $\n_{12},...,\n_{m1}$ for  the $r$-images of the facets of $\Star C$ at $r(C)$. The arrays of normal vectors
can also  be thought of as a map $\mathbf{n}:RimE(W_m) \cup RimV(W_m)
\rightarrow\R^n$.

 \textsf{Reduce-to-3D}  uses a subroutine \textsf{Is-Folded},  which checks for violations of the
 immersion assumption in cases where all normals are in the same plane. The output of \textsf{Reduce-to-3D} is either one of $\{0,1\}$ or a
realization of $W_m$ in $\R^3$, where the center of $W_m$ is mapped to
the origin. Output $1$ means that  the input $(W_m,\mathbf{n})$ is inconsistent with
our assumptions about the integrity of the input data or the immersion assumptions.
Output $0$ means that $\Star C$ has passed the local convexity check.

\vspace{-4mm}
\subsubsection{\emph{\textsf{Is-Folded}}}

Let $\vv,\uu,\w$ be three coplanar  non-zero vectors. The ordered triple ($\vv$, $\uu$,
$\w$) defines a plane angle at the origin in the following way: $\vv$ and $\w$ span
the two extreme rays of the angle, while $\uu$ is an interior vector of the angle -- i.e.,
the function of $\uu$ is to specify which of the two open subsets defined by $\vv$ and
$w$ is interior to the angle. We denote such angle by $\langle \vv|\uu|\w \rangle$. Note that $\langle \vv|\uu|\w \rangle=\langle \w|\uu|\vv \rangle$.

\textsf{Is-Folded} takes as input a 5-tuple of coplanar non-zero vectors
$(\aaa,\bb,\cc,\dd,\e)$, where $\aaa,\bb,\cc$ are pairwise distinct and
$\cc,\dd,\e$ are pairwise distinct. \textsf{Is-Folded} returns \emph{true} if the interiors of angles
$\langle \aaa|\bb|\cc \rangle$ and $\langle \cc|\dd|\e \rangle$ overlap and
\emph{false} otherwise.
For example, Figure 2 shows the case where ``folding" takes place: angle $\langle \cc|\dd|\e \rangle$ ``folds over" the angle $\langle \aaa|\bb|\cc \rangle$.
\begin{figure}[h]
\begin{center}
\resizebox{!}{140pt}{\includegraphics[clip=false,keepaspectratio=false]{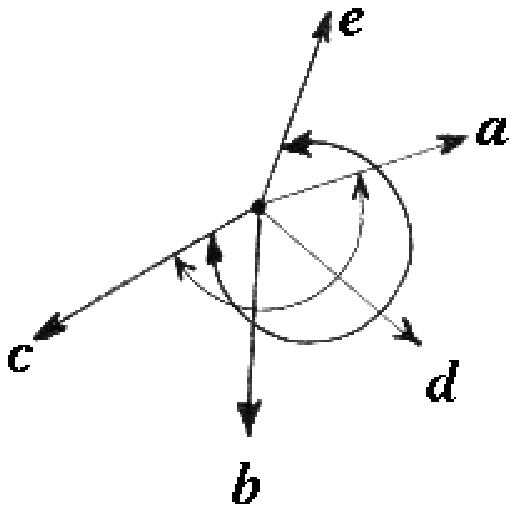}}
\caption{Angles $\langle \aaa|\bb|\cc \rangle$ and $\langle \cc|\dd|\e \rangle$ overlap.}
\end{center}
\end{figure}

In the pseudocode of this procedure we will use
 a boolean predicate $P(\vv|\uu_1,\uu_2|\w)$, which is defined for any 4-tuple of coplanar vectors
$\vv,\uu_1,\uu_2,\w$, where $\vv$ and $\w$ are distinct and $\uu_i  \neq  \vv$, $\uu_i
\neq \w$ for $i=1,2$. $P(\vv|\uu_1,\uu_2|\w)$ is \emph{false} if $\langle \vv|\uu_1|\w
\rangle \neq \langle \vv|\uu_2|\w \rangle$ (Figure 3, right) and \emph{true} otherwise (Figure 3, left).
\vspace{-12mm}
\begin{figure}[h]
\begin{center}
\resizebox{!}{200pt}{\includegraphics[clip=false,keepaspectratio=true]{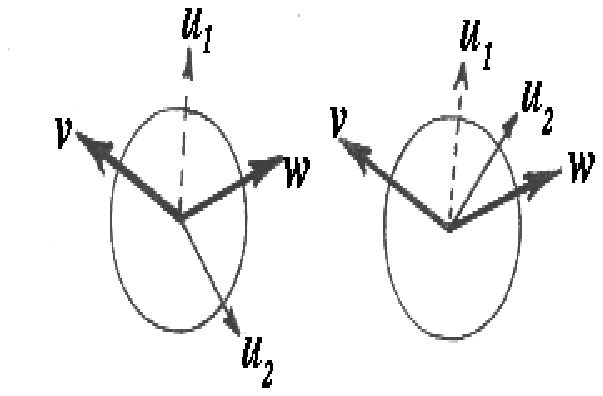}}
\caption{Left: $P(\vv|\uu_1,\uu_2|\w)=false$. Right:
$P(\vv|\uu_1,\uu_2|\w)=true$.}
\end{center}
\end{figure}
\vspace{-12mm}

\textbf{Input:} $\aaa,\bb,\cc,\dd,\e \in \R^n$, where $\dim \Span
\{\aaa,\bb,\cc,\dd,\e\}=2$,  $|\{\aaa,\bb,\cc\}|=3$, $|\{\cc,\dd,\e\}|=3$
\par
\textbf{Output:} boolean

\noindent \textbf{Algorithm} \textsf{Is-Folded}
\begin{algorithmic}
\If{$\aaa$ and $\e$ define the same ray} \If{$P(\cc|\bb,\dd|\aaa)=true$}
\Return{\emph{true}} \Else \quad \Return{\emph{false}} \EndIf  \EndIf

 \If{$P(\aaa|\bb,\e|\cc)=false$ {\bf and} $P(\cc|\dd,\e|\aaa)=true$ {\bf and} $P(\bb|\cc,\dd|\e)=true$}
\Return{\emph{false}} \Else \quad \Return{true}  \EndIf

\end{algorithmic}

 The following algorithm shows how to compute $P(\vv|\uu_1,\uu_2|\w)$ via standard linear algebra. For any ordered pair of vectors  $[\e_1,\e_2]$, such that  $\{\e_1,\e_2\}\subset \Span\{\vv,\uu_1,\uu_2,\w\}$, we use
$\sgn[\e_1,\e_2]$  to denote  the orientation of $[\e_1,\e_2]$ with respect to
some fixed orientation of $\Span\{\vv,\uu_1,\uu_2,\w\}$).
\pagebreak

\textbf{Algorithm} $P(\vv|\uu_1,\uu_2|\w)$
\begin{algorithmic}[h]
 \State $s \leftarrow \sgn[\vv,\w]$
 \If{$\sgn[\uu_1,\w]=\sgn[\vv,\uu_1]=s$} \If{$\sgn[\uu_2,\w]=\sgn[\vv,\uu_2]=s$} \Return{\emph{true}} \Else
 \quad
 \Return{\emph{false}} \EndIf \Else \If{$\sgn[\uu_2,\w]=\sgn[\vv,\uu_2]=s$} \Return{\emph{false}} \Else
 \quad
 \Return{\emph{true}} \EndIf \EndIf
\end{algorithmic}
\subsubsection{\protect Procedure \emph{\textsf{Reduce-to-3D}}}

\textbf{Input:} $W_m$: $m$-wheel graph,  $\mathbf{n}:RimE(W_m) \cup
RimV(W_m) \rightarrow \R^n$.   In the pseudocode we use $\n_i$ for $\mathbf{n}(i)$ and
$\n_{i\:i+1}$ for $\mathbf{n}(i\:i+1)$.
\par
\textbf{Output:} one of \{$0$; $1$; $(W_m,p)$ where $p:V(W_m) \rightarrow \R^3$\} )
\begin{algorithm}\label{alg:Reduce-to-3D}
\caption{ \textsf{Reduce-to-3D} }
\begin{algorithmic}[1]
\renewcommand{\algorithmiccomment}[1]{//#1}

\State $\e_1 \leftarrow \n_1$, $\bS \leftarrow \{\e_1\}$ \Comment{  $\bS$ is a maximal independent set of vectors in $(\overrightarrow{\aff r(C)})^{\perp}$}
\If{$\rank\{\n_1,\n_2\}=1$}

\If{$\n_2=\lambda \n_1$ for some $\lambda \ge 0$} \Return{1} \EndIf

\Else{ $\e_2 \leftarrow \n_2$, $\bS \leftarrow \bS \cup \{\e_2\}$}

\If{$m=2$} \If{\{$\rank\{\n_1,\n_{12},\n_{21}\}=2$\} {\bf and}
\{$\sgn[\n_1,\n_{12}]=\sgn[\n_1,\n_{21}]$\}} \Return{1} \Else \quad \Return{0} \EndIf
\EndIf

\EndIf

\State $i \leftarrow 3 \quad$  \Comment{rim-vertex counter}

\While{$i \neq 1 \mod m$ {\bf and } $|\bS|<3$}

\If{$\rank\{\e_1,\e_2,\n_i\} \le 2$} \quad \Comment{$\n_i$, $\e_1,\e_2$ are all in one
plane }

\If{$\n_i=\lambda \e_1$ {\bf or} $\n_i=\lambda \e_2$ for some $\lambda > 0$} \Return{1}

\EndIf

\If{$|\bS|=1$} \quad \Comment{$\e_1$ and $\e_2$ are collinear and contraoriented}

\If{$\sgn[\e_2,\n_{12}]=\sgn[\e_2,\n_{2m}]$} \Return{1} \Else \If{$m=3$} \Return{0} \Else{ $\e_2 \leftarrow \n_3$, $\bS \leftarrow \bS \cup \e_2$} \EndIf \EndIf

\Else \quad \Comment{in this case we know  $|\bS|=2$}

\If{$\textsf{Is-Folded}(\e_1,\n_{12},\e_2,\n_{i-1\:i},\n_i)=true$} \Return{1} \EndIf

\EndIf

\Else{ $\e_3 \leftarrow \n_i$,  $\bS \leftarrow  \{\e_3\}$}

\EndIf

\State{$i \leftarrow i+1$}

\EndWhile

\If{$i=1 \mod m$ {\bf or} $|\bS|<3$} \Return{0} \EndIf

\State{ $\fp(0) \leftarrow (0,0,0)$}

\For{$j=1$ to $m$} \State{ $\fp(j) \leftarrow (\e_1 \cdot \n_j, \e_2 \cdot \n_j, \e_3
\cdot \n_j)$} \EndFor \State\Return{$(W_m,\fp)$}

\end{algorithmic} \end{algorithm}
\clearpage

\subsection{Procedure \emph{\textsf{Check-if-Cone-Convex}}}
\vspace{-8mm}
Denote by $\fp[W_m]$ the simplicial surface that results from the map $\fp:V(W_m) \rightarrow \R^3$ (see Subsection 6.3).
Informally speaking, we test the surface $\fp[W_m]$ for convexity by going
around the wheel and checking whether the following two conditions are satisfied or not. The
first condition will be the conical PL-surface $\fp[W_m]$ "turning in the same
direction" each time we increment the index: here, the formal meaning of ``the same direction" is captured by the
linear-algebraic notion of orientation of a frame. Recall that the sign of a list
$[\vv_1,\vv_2,\vv_3]$ of  three vectors in $\R^3$ is  the sign of the determinant of the $3
\times 3$ matrix whose $i$-th row is $\vv_i$, which we denote by $\sgn
[\vv_1,\vv_2,\vv_3]$.  If $\vv_1,\vv_2,\vv_3\in \R^n$ with $n>3$,  then to define the sign of the triple $[\vv_1,\vv_2,\vv_3]$ we need to fix an orientation in a 3-subspace of $\R^n$ containing $\{\vv_1,\vv_2,\vv_3\}$.
Let $\sgn [\fp(1),\fp(2),\fp(3)]=s \neq 0$. Formally, the first  condition is that for every  $i \in V(W_m)$ the sign $\sgn [\fp(i),\fp(i+1),\fp(i+2)]$ must be $s$ or $0$; the zero sign corresponds to the case where the surface "continues straight", i.e., vectors
$\fp(i),\fp(i+1),\fp(i+2)$ lie in one plane. The second condition will be $\fp[W_m]$
not intersecting itself. This can be captured by checking the sign of every triple
of the form $[\fp(1),\fp(i),\fp(i+1)]$ is $0$ or $s=\sgn [\fp(1),\fp(2),\fp(3)]$. Thus,
roughly speaking, the first condition ensures that the convexity is not lost due to a
turn in the wrong direction, while the second condition guarantees the cone will
not intersect itself. Finally, the correctness of  \textsf{Check-if-Cone-Convex}
hinges on the following lemma, whose proof we omit.
\vspace{-8mm}
\begin{lemma} Let $\fp : V(W_m) \rightarrow \R^3$ be a  realization of the $m$-wheel graph ($m\ge3$) which does not map any two consecutive spokes into the same ray. Suppose $\sgn [\fp(1),\fp(2),\fp(3)] \neq 0$. Then $\fp[W_m]$  lies on the boundary of a convex cone if and only if for all $0 \le i \le m-1$\emph{:}

 1) $\sgn [\fp(i),\fp(i+1),\fp(i+2)]=\sgn
[\fp(1),\fp(2),\fp(3)]$ and

2) $\sgn [\fp(1),\fp(i+1),\fp(i+2)]=\sgn [\fp(1),\fp(2),\fp(3)].$
\end{lemma}
\vspace{-2mm}
 The input to \textsf{Check-if-Cone-Convex-in-3D} is  wheel graph $W_m$
$(m>1)$, each of whose vertices $v$ is equipped with a corresponding point $\fp(v)$ in
$\R^3$. For notational simplicity we assume that
the vertex set $V(W_m)$ of $W_m$ is $\{0,1\ldots,m\}$, where $0$ is the center of
$W_m$, and $[1,\ldots m]$ is a cyclic order on the rim vertices. In addition, we assume that \textit{vertex
$0$ is realized at the origin $\mathbf{0}$}.

\textbf{Input:} $W_m$: wheel
graph, $\fp:V(W_m)\rightarrow \R^3$
\par
\textbf{Output:} boolean
\pagebreak

\begin{algorithm} \caption{\textsf{Check-if-Cone-Convex}} \label{alg:cone-check}
\begin{algorithmic}
 \State $sign \leftarrow \sgn [\e_1,\e_2,\e_3]$ \For{$j$ from $2$ to $m$} \If{\{$\sgn
[\fp(j),\fp(j+1),\fp(j+2)]=-sign$\} \textrm{\bf{or}} \{$\sgn [\e_1,\e_2,\fp(j+2)]=-sign$\}}
     \State \textbf{return} \textit{false}
\EndIf \EndFor \State \textbf{return} \textit{true} \end{algorithmic} \end{algorithm}
\vspace{-2mm}
\section{Complexity Analysis }\label{sec:Complexity}
\vspace{-2mm}
\par \noindent In this section  $\ssf_k$ denotes the number of $k$-cells of $(\cM,\mathcal{P})$, and $\ssf_{kl}$  the number of
incidences between $k$-cells and $l$-cells of $(\cM,\mathcal{P})$; if $\mathcal{P}'$ is a subposet
of $\mathcal{P}$, then $\ssf_k(\mathcal P')$ denotes the number of $k$-cells in $\mathcal{P}'$. Note that
all linear algebra in the algorithm is essentially reduced to comparisons of signs of lists of at
most three $n$-vectors; we will refer to any such calculation as a \emph{sign
computation.} Unless mentioned otherwise, as e.g. in the next Subsection, we assume that the input is in the standard form.

\begin{enumerate}
 \item Building the wheel graph for $\Star C$ takes time linear
in the number of ridges  of $\Star C$.
  \item Since \textit{Corners} is accessed at most $\ssf_{n-3}$ times, \textsf{Reduce-to-3D}
   is called at most $\ssf_{n-3}$ times.
  \item $\textsf{Reduce-to-3D}(W_m,\n)$ requires at most $O(m)$ sign computations.
  \item \textsf{Check-if-Cone-Convex} requires at most $O(m)$ sign computations.
  \item  \textsf{Is-Folded} requires a constant number of sign computations.
\end{enumerate}
\begin{description}

\item[1]\emph{Suppose the algorithm uses the  field arithmetic ($+,-,\times,\div$) and each arithmetic operation has unit
cost. } This model is realistic when real computations are conducted with floating
point arithmetic. If  $n$ is fixed, the complexity of the algorithm is
$O(\ssf_{n-3\:n-2})=O(\ssf_{n-3\:n-1})$. To estimate the complexity in the case where $n$ is one of the parameters describing the input size,  we need to estimate the contributions of \emph{sign computations} in (3)--(5). Notice that any sign computation in (3)-(5) deals with,
at most, six $n$-vectors. Since standard linear-algebraic procedures over a field can be
used,  the complexity of the algorithm is $O(n\ssf_{n-3\:n-2})$.

\medskip
\item[2] \emph{What follows is a discussion of  the complexity in the cases where no floating point error can be tolerated.} Let $\cR$   be
the base ring of the computational model: i.e., all numerical input data (such as  the
coordinates of vertices, the coefficients of normals to $(n-1)$-faces etc.) come from
$\cR$. Furthermore, we assume that $\Z \le \cR \le \R$ in the partial order of rings.
When we discuss the degrees of the polynomial predicates evaluated by the
algorithm, we consider them as polynomials with integer coefficients in the input
parameters. In this context the phrase  \emph{arithmetic operation} stands for any ring-theoretic
operation ($+,-,\times$).

Case 1: \emph{the dimension $n$ is fixed.} In this case all linear-algebraic
computations can be done via determinants. Using determinants  has an advantage of keeping the degrees
of evaluated polynomial predicates at bay. Moreover, since in our algorithm the largest determinants
are $3\times3$, the highest degree of evaluated predicates is 3. Thus, the
arithmetic complexity of the algorithm is $O(\ssf_{n-3\:n-2})$ and the algorithm
evaluates at most $O(\ssf_{n-3\:n-2})$ polynomial predicates of degree 3.

Case 2: \emph{the dimension $n$ is not fixed.} If $n$ is not too large, the linear-algebraic
computations can still be done via direct determinant evaluations. In each computation  we are
dealing with at most three $n$-vectors, which means that we may have to evaluate $\binom{n}{n-3}$ $3\times3$
determinants to find a minor of maximal rank. Thus, the total arithmetic complexity of the algorithm is
$O(n^3\:\ssf_{n-3\:n-2})$. The case of large values of $n$ is considered in the next
paragraph.
\end{description}
\vspace{-4mm}
\subsection{Exact  Computations over $\Z$.}
\vspace{-4mm}
In here we consider the case of exact computation. The dimension $n$ is not fixed and $\cR=\Z$.
 We are now interested in the bit complexity of our algorithm, e.g., in
the multitape Turing machine model. Note that the same techniques can be used for sign computations in our algorithm. Since each sign computation involves no more than 6 vectors, the bit complexity of each sign computation using Yap's (2002) ramification of  the Bachem-Kannan algorithm is $O(n\textsf{M}_b(L))$, where $\textsf{M}_b(x)$ is the bit-complexity of multiplication of two integers of binary sizes not exceeding $x$ and $L$ is a bound on the binary size of the coefficients of the vectors (see (Yap, 2002) for details).
Then the total complexity of the algorithm is $O(n\ssf_{n-3\:n-2}\textsf{M}_b(L))$. Devillers
et al. (1998) have shown  that any convexity checker, \emph{whose work  does not depend on
the nature of {\slshape\textsf{R}}}, has to evaluate at least one polynomial  of
degree $n$ -- however, this lower bound is mandatory only for those checkers that work
the same way for any $\cR$.
\begin{theorem}Let $r:\cM \rightarrow \R^n$ be a cellwise-flat PL-realization of a
manifold $(\cM,\mathcal{P})$ of dimension $n-1$.  Suppose the \emph{input is in the standard
form} and all normals have integer coordinates of binary size not exceeding $L$. There
exists a polynomial time algorithm  for checking convexity of  $r:\cM \rightarrow \R^n$  with (multitape Turing machine) complexity  of \emph{$O(n\ssf_{n-3\:n-2}\textsf{M}_b(L))$}.
\end{theorem}

Now, let us consider the situation where the input is given in the traditional form, i.e. as
the poset $\mathcal{P}[0,n-3,n-2,n-1]$, equipped
with the coordinates of the vertices. If we  have no restrictions on the combinatorics and geometry of the geometric realization of cells of $(\cP,\cM)$, then it is very difficult, or even impossible, to construct Euclidean inner normals from given data. Let us assume that the partition $\mathcal{P}$
is simplicial.  In order to do sign computations, we need first write down
\emph{Euclidean} inner normals for all corner-facet and
corner-ridge incidences. For each such incidence we have to deal with roughly  $n$
vectors of length $n$. Computing a Euclidean normal is then reduced to a fidning a non-zero solution for a homogeneous system $M\x=0$
where $M$ is at most $n$ by $n$ matrix. We can use Yap's (2002) version of
the Bachem-Kannan algorithm to compute (upper triangular) Hermit's Normal Form for
the system $M\x=0$. Then a non-zero solution vector of at most polynomial size  can be found  in polynomial time by using standard techniques of linear algebra: we just work our way  from the bottom of the normalized matrix up until all $x_i$'s are found. Alternatively, one can use a polynomial  algorithm in Yap (2002: Sec. 10.8-10.9), based on repeated application of Bachem-Kannan algorithm, to further reduce the system to Smith Normal Form and then find a solution. Furthermore, to reduce the complexity, we can deal with each corner $C$ in the following way. If $\vv_0,...,\vv_n$ are the vertices of $r(C)$, then we first find  Hermit's normal form for the matrix $[\vv_1-\vv_0,...,\vv_n-\vv_0]$ and then, for each $P$, where $P$ is a  ridge or a facet incident to $C$, compute
an integral Euclidean normal vector to $r(P)$ at $r(C)$. Then the complexity of all linear-algebraic computations for $\Star C$ is dominated by the complexity of finding  Hermit's normal form for the matrix $[\vv_1-\vv_0,...,\vv_n-\vv_0]$, which is $O(n^3\textsf{M}_b(L))$ (Yap, 2002). Thus , the total complexity  is $O(n^3\ssf_{n-3\:n-2}\textsf{M}_b(L))$.

If $n=3$, then the corners are the vertices and the required normals are easy to produce. Now, suppose $n>3$. How large the coefficients of the integral Euclidean normals, discussed above, can be? It is obviously possible to produce each such Euclidean normal as a vector whose coordinates are polynomials of degree at most $n-3$ in the
coordinates of the vertices. Furthermore, Siegel \citep[see e.g.][]{Yap}
 proved that a homogeneous system of
$k$ linear equations with $n$ variables over $\Z$ has a non-zero solution where each
component is bounded in absolute value by $1+(nA)^{\frac{k}{n-k}}$ (for us $k=n-3$)
where $A$ is the largest of the absolute values of the coefficients. Siegel also
showed this bound could not be improved. When $n$ is small enough, a vector satisfying Siegel's bound can be
found by classical methods of lattice reduction (no efficient methods for finding such a vector are known for large $n$).
If normals satsifying Siegel's bound are used in the algorithmic precedures given above, then  the largest integers that may appear in sign computations via determinants are of the order
$\lambda (nA_0)^{n-3}$ -- where $\lambda \gtrapprox 1$ is a constant slightly larger than
1, and $A_0$ is twice the largest of the absolute values of the vertex coordinates.

\begin{theorem}Let $r:\cM \rightarrow \R^n$ be a cellwise-flat PL-realization of a \emph{simplicial}
manifold $(\cM,\mathcal{P})$ of dimension $n-1$.   Suppose the input consists of the poset
$\mathcal{P}[0,n-3,n-2,n-1]$ equipped with the coordinates of the vertices, and that for each vertex $\vv=r(v)$ of $r(\cM)$ we have $\vv \in \Z^n$ and $|\vv|\le 2^L$. There
exists a polynomial time algorithm  for checking convexity of  $r:\cM \rightarrow \R^n$  with (multitape Turing machine) complexity  of \emph{$O(n^3\ssf_{n-3\:n-2}\textsf{M}_b(L))$}.
\end{theorem}

The input requirements in the above theorem can be relaxed. If we know only
$\mathcal{P}'=\mathcal{P}[0,n-3,n-2]$ (or $\mathcal{P}'=\mathcal{P}[0,n-3,n-1]$)
together with the circular order of facets (or ridges) at all ridges, then
 $\mathcal{P}[0,n-3,n-2,n-1]$
 can be computed at no extra cost.


\vspace{-4mm}
\subsection{Surfaces in $\R^3$}\label{sec:n=3}
\vspace{-4mm}
The algorithm runs in linear time in the number of vertices  when $\cM$ is spherical. However, a sequence of non-spherical
PL-manifolds of dimension 2 can have the edge number  growing quadratically in
$\ssf_0$. Thus, it is desirable to check the topological type of the input by just
counting 1-cells (edges) in  $\mathcal{P}[0,1]$: once their number exceeds $c\ssf_0$,
where $c$ is some constant which is easy to calculate,  we stop and declare the input
non-convex. This check helps preserve the $O(\ssf_0)$ running time bound for PL-surfaces in $\R^3$. One may
wonder if such a check is necessary, as it seems  very likely our algorithm will quickly
encounter a non-convex vertex, if the input surface  is homeomorphic to a sphere-with-handles or
sphere-with-M\"{o}buis-strips. Surprisingly, Betke \& Gritzmann (1984), proved that
any orientable non-spherical connected closed 2-manifold can be PL-embedded into
$\R^3$ so that it has exactly 5 non-convex vertices but no fewer! The problem of determining the minimal possible number of non-convex vertices in a PL-immersion of a non-orientable closed 2-manifold is open.

In the case of  $\R^{3}$ the requirements on the combinatorial part of the input can be somewhat
relaxed: in what follows we show it is sufficient to  know  only $\mathcal{P}[0,1]$, which is the 1-skeleton graph of
$(\cM,\cP)$. First, the planarity of this graph can be checked in $O(\ssf_0)$ time (Lempel et al, 1967). For a planar graph we can also determine the faces in linear time -- i.e., in $O(\ssf_0)$ time we can create the face-nodes, where each face-node is double-linked to its edge-nodes.  Once we know the faces in terms of their edges, we can double link each face-node to the
vertices-nodes of all of its edges. Because of the sphericity of $(\cM,\mathcal{P})$ the latter
task takes $O(\ssf_0)$ time. Thus, the adjacency list representing $\mathcal{P}[0,1,2]$
can be constructed from  the adjacency list representing $\mathcal{P}[0,1]$ in $O(\ssf_0)$
time.

The case of  $\R^3$ is a rather special one. First, $n=3$ is the smallest dimension for which the techniques of this paper apply.
Second, even in the case of $\R^n$ the convexity test for each corner is reduced to  testing convexity of a section of the star of this corner, which is essentially equivalent to testing convexity of a cone in $\R^3$. Another important consideration is that in application a 2-surface in $\R^3$ is normally specified by its combinatorics and the coordinates of the vertices or equations for the facets: it is therefore important to specify how our algorithm can be applied when the imput is given in the traditional form. Namely, suppose we are given $\mathcal{P}[0,1,2]$ equipped with the coordinates of the vertices
$\vv_1=r(v_1),\ldots,\vv_{\ssf_0}=r(v_{\ssf_0})$. The corner-ridge normals are then just vectors $\vv_i-\vv_j$. The question remains how to find corner-facet normals, i.e. vectors pointing from the vertices of the facets into the interiors of the facets. This is easy if it is known that the facets are convex.
Otherwise we have the following algorithmic problem. Let $C_k$ be the $k$-cycle graph. Consider a rectilinear embedding $r$  of $C_n$  in an affine plane $A \subset \R^3$  -- the pair $(C_k,r)$ defines a 2-dimensional polytope  $P(C_k,r)$ whose boundary is $r(C_k)$ (here $C_k$ is regarded as PL-manifold).  Let $v$ be a vertex of $C_k$.  The problem is to find a non-zero vector $\n \in \overrightarrow{A}$ such that $r(v)+\epsilon\n$ lies in the interior of $P(C_k,r)$. This problem can be solved in time $O(k)$; solving this problem for all facets will require $O(\ssf_0)$ ring-arithemtic operations. Thus, there is no difference in time-complexity betwen the standard and traditional forms of the input for $n=3$.
We will now restate the observations made in this Section in the following theorem.

\begin{theorem}Let $r:\cM \rightarrow \R^3$ be a cellwise-flat PL-realization of a
2-manifold $(\cM,\mathcal{P})$.  Suppose we are given the 1-skeleton of $(\cM,\mathcal{P})$ equipped with the coordinates of the vertices. Let $L$ be the upper bound on the bit sizes of the coordinates of the vertices. There
exists an algorithm  for checking convexity of  $r:\cM \rightarrow \R^n$  with (multitape Turing machine) complexity  of \emph{$O(\ssf_{0}\textsf{M}_b(L))$}.
\end{theorem}
\vspace{-4mm}
\section{Conclusions}
\vspace{-4mm}
This paper describes a local approach to convexity verification of PL-hypersurfaces. The main theoretical result of the paper is a characterization of global convexity of a PL-hypersurface in $\R^n$  in terms of the local convexity properties of the surface at its $(n-3)$-faces. Building on this approach we give a polynomial-time convexity checking algorithm that can be applied  for any closed hypersurface. The approach presented in this paper can be generalized to piecewise-polynomial surfaces of small degree.

\end{document}